\UseRawInputEncoding
\documentclass[twocolumn,showpacs,superscriptaddress,amsmath,amssymb,aps,pra]{revtex4-1}
\usepackage{bm}
\usepackage{graphicx}
\usepackage{dcolumn}
\usepackage{amsmath,txfonts}
\usepackage{epstopdf}
\usepackage[mathlines]{lineno}
\usepackage{color}
\usepackage[colorlinks=true,linkcolor=blue,citecolor=blue]{hyperref}
\usepackage[normalem]{ulem}

\begin{document}

\title{Generating Bell states and $N$-partite $W$ states of long-distance qubits in superconducting waveguide QED}

\author{Guo-Qiang Zhang}
\affiliation{School of Physics, Hangzhou Normal University, Hangzhou, Zhejiang 311121, China}

\author{Wei Feng}
\affiliation{School of Physics, Hangzhou Normal University, Hangzhou, Zhejiang 311121, China}

\author{Wei Xiong}
\affiliation{Department of Physics, Wenzhou University, Zhejiang 325035, China}

\author{Da Xu}
\affiliation{Interdisciplinary Center of Quantum Information, State Key Laboratory of Modern Optical\\
Instrumentation and Zhejiang Province Key Laboratory of Quantum Technology and Device,\\
School of Physics, Zhejiang University, Hangzhou 310027, China}

\author{Qi-Ping Su}
\affiliation{School of Physics, Hangzhou Normal University, Hangzhou, Zhejiang 311121, China}

\author{Chui-Ping Yang}
\email{yangcp@hznu.edu.cn}
\affiliation{School of Physics, Hangzhou Normal University, Hangzhou, Zhejiang 311121, China}

\begin{abstract}
We show how to generate Bell states and $N$-partite $W$ states of long-distance superconducting (SC) qubits in a SC waveguide quantum electrodynamical (QED) system, where SC qubits are coupled to an open microwave transmission line. In the two-qubit case, the Bell state of two long-distance qubits can be a dark state of the system by choosing appropriate system parameters. If one proper microwave pulse drives one of two qubits, the two qubits will evolve from their ground states to a Bell state. Further, we extend this scheme to the multi-qubit case. We show that $W$ states of $N$ long-distance qubits can also be generated. Because both the Bell and $W$ states are decoupled from the waveguide (i.e., dark states of the system), they are steady and have very long lifetimes in the ideal case without decoherence of qubits. In contrast to the ideal case, the presence of decoherence of qubits limits the lifetimes of the Bell and $W$ states. Our study provides a novel scheme for generating Bell states and $N$-partite $W$ states in SC waveguide QED, which can be used to entangle long-distance nodes in waveguide quantum networks.
\end{abstract}

\date{\today}

\maketitle

\section{Introduction}

Due to its fundamental importance for showing quantum nonlocality and diverse applications in quantum technologies, quantum entanglement has attracted substantial interest in the last decades~\cite{Horodecki09,Aolita15}. The Bell state refers to the maximally bipartite entangled state~\cite{Kwiat95,Quiroga99}. Correspondingly, there are two important classes of multipartite entangled states, i.e., Greenberger-Horne-Zeilinger (GHZ) state~\cite{Greenberger90} and $W$ state~\cite{Dur62}, where the latter is more robust against the loss of excitation~\cite{Brunner14}.
Besides Bell states, GHZ states and $W$ states, the NOON states~\cite{Sanders89} and Affleck-Kennedy-Lieb-Tasaki (AKLT) states~\cite{Affleck87} are two important types of entangled states.
Compared to the Bell state, the $W$ state is more favourable in quantum information processing, because it entangles more qubits. To date, Bell states, GHZ states, and $W$ states have been widely studied in, e.g., cavity quantum electrodynamical (QED) systems~\cite{Zheng01,Guo02,Deng06}, cavity-magnon systems~\cite{Yuan20,Qi22}, superconducting (SC) circuit systems~\cite{Ansmann09,Neeley10,Yang11,Yang12,Yang13,Li19,Stojanovic20,Zhang20,Peng21,Feng22}, and cold neutral atoms~\cite{Stojanovic21}, where the entangled qubits are short-distance. However, in practical applications, it is vital to entangle distant nodes in a quantum network~\cite{Ribordy00,Simon03,Venuti06,He09,Zou21,Mok20,Hu21,Zhang23,Brask10,Trifunovic13,Krutyanskiy23}. Therefore, generating Bell states, GHZ states, and $W$ states of long-distance qubits is urgent and necessary~\cite{Zou02,Eibl04,Zhou23,Kim20,Jiang23}.

The waveguide QED provides an excellent platform for generating long-distance entanglement~\cite{Tudela11,Zheng13,Ballestero15,Liao15,Facchi16,Mirza16,Yang21}, where distant atoms, both natural and artificial, interact with the continuous traveling modes in a one-dimensional (1D) waveguide~\cite{Roy17,Gu17}. Experimentally, the waveguide QED has been implemented in quantum dots coupled to a metallic nanowire~\cite{Akimov07}, SC qubits coupled to an open microwave transmission line~\cite{Astafiev10,Hoi12,Hoi15}, natural atoms coupled to an optical fiber~\cite{Bajcsy09,Goban12,Sorensen16,Corzo16}, nanoparticles coupled to a silica nanofiber~\cite{Petersen14}, and so on. Compared with other waveguide QED systems, the SC waveguide QED system has its unique merits, such as the achievable strong coupling (even ultrastrong coupling) of SC qubit to the open transmission line, the small dissipation of SC qubit, good scalability, and easy controllability~\cite{Roy17,Gu17}. Based on SC waveguide QED systems, many exotic phenomena have been explored, such as resonance fluorescence~\cite{Astafiev10}, giant SC atoms~\cite{Kockum18,Kannan20,Vadiraj21}, collective Lamb shifts~\cite{Wen19}, and three-state dressed states~\cite{Koshino13}. Here, the giant SC atom means that an artificial SC atom is coupled to a SC waveguide at several points, where the dipole approximation is invalid because the distance between different points is comparable to (or larger than) the wavelengths of travelling microwave modes in the waveguide.
In optical waveguide QED systems, the Bell~\cite{Zhang19,Zhan22} and four-partite $W$ states~\cite{Song22} can be prepared with the assistance of photon counting detection or homodyne detection. Very recently, Ref.~\cite{Santos23} has proposed to generate the Bell state with giant SC atoms in a SC waveguide.

\begin{figure}
\includegraphics[width=0.48\textwidth]{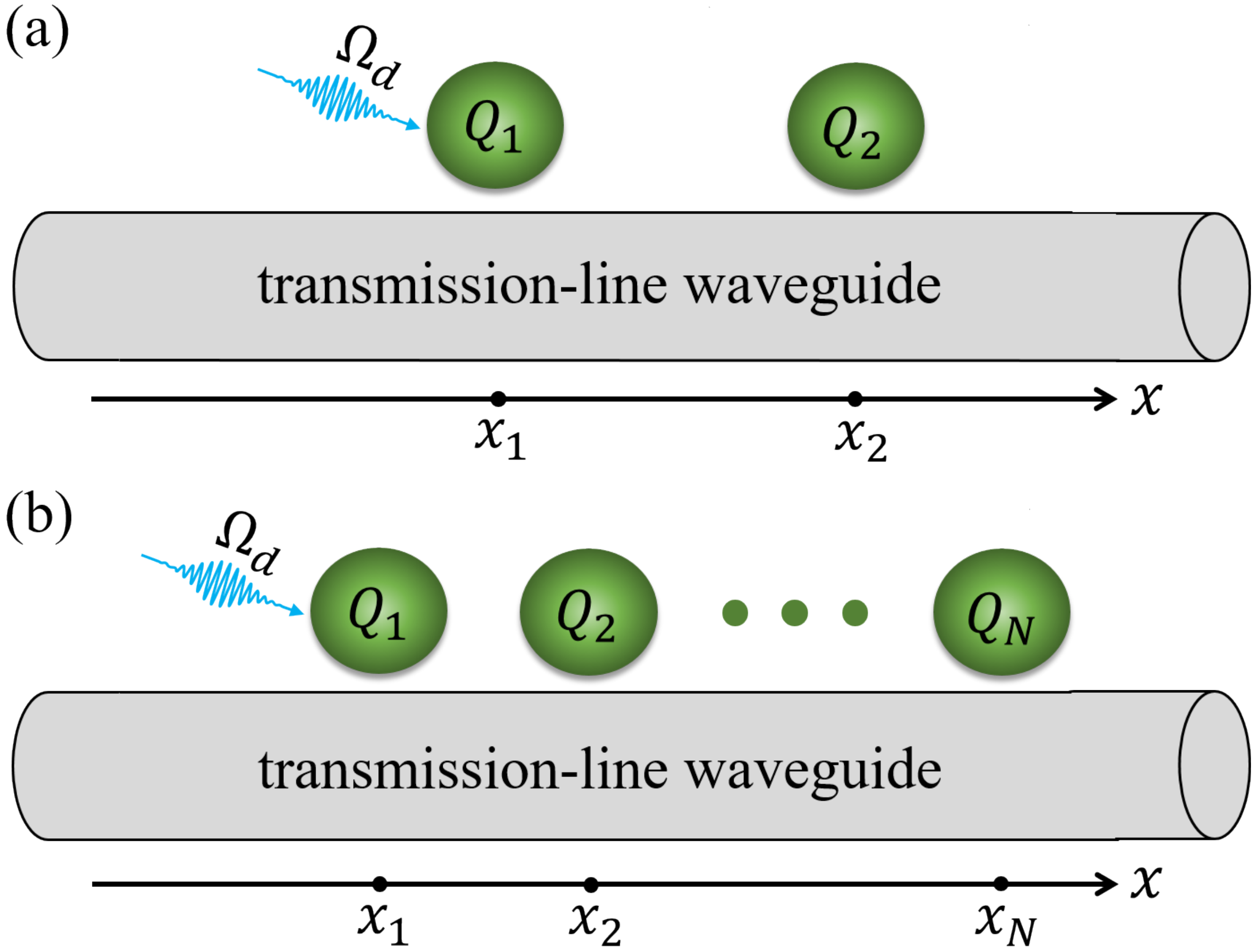}
\caption{Schematic diagrams of the considered waveguide QED systems. (a) Two transmon qubits and (b) $N$ ($\geq 3$) transmon qubits are coupled to a 1D SC transmission line, where the location of the $j$th qubit $Q_j$ is denoted as $x_j$. To generate the Bell and $N$-partite $W$ states, one microwave pulse with Rabi frequency $\Omega_d$ drives the qubit $Q_1$.}
\label{fig1}
\end{figure}

In this paper, we propose a scheme for generating the Bell and $N$-partite $W$ states with long-distance SC qubits in a SC waveguide without need of any measurements. First, we consider two identical SC transmon qubits, $Q_1$ and $Q_2$, coupled to a 1D SC transmission line [see Fig.~\ref{fig1}(a)]. Under the Born-Markovian approximation, the considered SC waveguide QED system can be described by a Lindblad master equation~\cite{Lalumiere13}. With proper system parameters, we find that the Bell state of $Q_1$ and $Q_2$ is exactly a dark state of the waveguide QED system. By driving the qubit $Q_1$ with a proper microwave pulse, the two-qubit ground state will evolve into a dark state (i.e., the Bell state). Further, the proposed scheme can be easily extended to the multi-qubit case. In a SC waveguide QED system with $N$ ($N\geq 3$) long-distance transmon qubits [see Fig.~\ref{fig1}(b)], the $N$-partite $W$ state can also be a dark state of the system by properly choosing the coupling strength of each qubit to the waveguide. Under the drive of an appropriate microwave pulse, the $N$-partite $W$ state of long-distance qubits can also be generated. In the absence of the intrinsic decoherence of qubits, the generated Bell and $N$-partite $W$ states are steady because they are decoupled from the SC waveguide. However, in realistic experimental conditions, the decoherence from the qubits determines the lifetimes of the Bell and $N$-partite $W$ states.

Note that our work is different from Ref.~\cite{Santos23} in the following two aspect. First, Ref.~\cite{Santos23} is for the generation of Bell states, while our study discusses how to prepare not only Bell states but also $N$-partite $W$ states. Second, our scheme is based on transmon qubits instead of giant SC atoms used in Ref.~\cite{Santos23}. As compared with giant SC atoms, transmon qubits are commonly used and easily fabricated in SC-circuit experiments~\cite{Gu17,Kannan20}. Because no measurement is required and the initial state is the ground state of the waveguide QED system, our scheme can be easily performed in experiments. The generated Bell and $N$-partite $W$ states in waveguide QED have potential applications in constructing waveguide quantum networks~\cite{Zhang19,Zhan22,Song22,Santos23}.

The paper is organized as follows. In Sec.~\ref{model}, we derive the effective non-Hermitian Hamiltonian of the SC waveguide QED system from the Lindblad master equation. Using the obtained non-Hermitian Hamiltonian, we analyze the eigenvectors and eigenvalues of the waveguide QED system. In Sec.~\ref{Generation-Bell}, we present the scheme for generating Bell states of two long-distance qubits and explain the related physical mechanism using the system eigenvectors. In Sec.~\ref{Generation-W}, we extend the scheme to the multi-qubit case and show how to prepare $W$ states of $N$ long-distance qubits. Our discussions and conclusions are given in Sec.~\ref{summary}.
In addition, three appendices are included. In Appendix~\ref{Appendix-A}, we derive the master equation of the waveguide QED
system with transmon qubits. In Appendix~\ref{Appendix-B}, we give the parameters used in numerical simulations. In Appendix~\ref{Appendix-C}, we discuss the effects of the position deviations and the higher levels of transmon qubits on our scheme.

\section{Model}\label{model}

As illustrated in Fig.~\ref{fig1}, $N$ ($\geq 2$) identical transmon qubits, $Q_j$ located at $x_j$ ($j=1,2,\cdots,N$), are coupled to a 1D SC transmission line. When $Q_1$ is driven by a microwave pulse with Rabi frequency $\Omega_d$, the total Hamiltonian of the $N$ transmon qubits is (hereafter assuming $\hbar=1$)
\begin{equation}\label{original-Hamiltonian}
H=\omega_q\sum_{j=1}^N \sigma_j^+\sigma_j^- + \Theta (t_0-t)\Omega_d(\sigma_1^+ e^{-i\omega_d t}+\sigma_1^- e^{i\omega_d t}),
\end{equation}
where $\omega_q$ is the $|0_j\rangle\leftrightarrow|1_j\rangle$ transition frequency of $Q_j$ with the ground state $|0_j\rangle$ and the excited state $|1_j\rangle$, $\sigma_j^+=|1_j\rangle\langle0_j|$ and $\sigma_j^-=|0_j\rangle\langle1_j|$ are the ladder operators of $Q_j$, $t_0$ is the duration of the drive pulse, $\omega_d$ is the frequency of the drive pulse, and $\Theta (t_0-t)$ is the Heaviside function.

By taking trace over the degrees of freedom of the waveguide (i.e., the SC transmission line) at zero temperature, we find that the density operator $\rho$ of $N$ qubits satisfies the following Born-Markovian master equation~(see Appendix~\ref{Appendix-A})~\cite{Lalumiere13}:
\begin{eqnarray}\label{master-1}
\frac{\partial \rho}{\partial t}&=&-i[H,\rho]+\sum_{j=1}^{N}\sum_{m=1}^{N}\frac{\kappa_{jm}}{2}\mathcal{D}[\sigma_j^-,\sigma_m^+]\rho\nonumber\\
                                & &+\frac{\gamma}{2}\sum_{j=1}^N\mathcal{D}[\sigma_j^-,\sigma_j^+]\rho
                                   +\frac{\gamma_\varphi}{2}\sum_{j=1}^N (\sigma^{\,z}_j \rho\sigma^{\,z}_j - \rho),
\end{eqnarray}
with the Lindblad superoperator
\begin{equation}\label{}
 \mathcal{D}[\sigma_j^-,\sigma_m^+]\rho=2\sigma_j^-\rho\sigma_{m}^+ -\sigma_{m}^+\sigma_j^-\rho
                                                 -\rho\sigma_{m}^+\sigma_j^-.
\end{equation}
Note that Eq.~(\ref{master-1}) is only valid when the location of $Q_j$ is given by $x_j=\pm l\pi/k$ ($l=0,1,2,\cdots$), where $k=\omega_q/\upsilon$ is the wave vector, and $\upsilon$ is the speed of the microwave at frequency $\omega_q$ in the waveguide. Here the second term in Eq.~(\ref{master-1}) presents both the cooperative dissipations and the local dissipations of $N$ qubits caused by the waveguide, where $\kappa_{jm}=2c_jc_{m}\kappa$ is the cooperative decay rate of $Q_j$ and $Q_m$ for $j\neq m$ (local decay rate of $Q_j$ for $j= m$), $\kappa=2\pi g^2\omega_q$ is the collective decay rate of $N$ qubits, $c_j=(g_j/g)e^{-ikx_j}$ characterizes the relative coupling strength of $Q_j$ to the waveguide, and $g$ ($g_j$) is the coupling strength between $N$ qubits ($Q_j$) and the waveguide. In addition, the two terms in the second line of Eq.~(\ref{master-1}) denote the intrinsic decoherence of $N$ qubits, where $\gamma$ ($\gamma_\varphi$) is the energy relaxation rate (pure dephasing rate) of the individual qubit, and $\sigma_j^{\,z}=|1_j\rangle\langle1_j|-|0_j\rangle\langle0_j|$ is the Pauli operator related to $Q_j$.

In the absence of both the drive pulse and the intrinsic decoherence of $N$ qubits (corresponding to the case with $\Omega_d=\gamma=\gamma_\varphi=0$), the master equation in Eq.~(\ref{master-1}) can be rewritten as~\cite{Minganti19,Zhang21,Chen21}:
\begin{eqnarray}
\dot{\rho} =-i\left(H_{\rm eff}\rho-\rho H_{\rm eff}^\dag\right)+\sum_{jm}\kappa_{jm}\sigma_j^-\rho\sigma_{m}^+,
\end{eqnarray}
where $H_{\rm eff}$ is the effective non-Hermitian Hamiltonian of the waveguide QED system given by
\begin{eqnarray}\label{Hamiltonian-eff}
H_{\rm eff}&=&\sum_{j=1}^{N}\left(\omega_q-i\frac{\kappa_{jj}}{2}\right) \sigma_j^+\sigma_j^-
              -i\sum_{j=1}^{N}\sum_{m=j+1}^{N}\frac{\kappa_{jm}}{2}(\sigma_{j}^+\sigma_m^- + \sigma_j^-\sigma_{m}^+).\nonumber\\
\end{eqnarray}
For simplicity and clarity, we have ignored the intrinsic decoherence of $N$ qubits in deriving the above effective Hamiltonian. Due to $\gamma,\gamma_\varphi \ll\kappa_{jm}$, it is enough to explain the related physical mechanism of preparing Bell states and $W$ states via using the eigenvectors of $H_{\rm eff}$ (cf.~Secs.~\ref{Generation-Bell} and \ref{Generation-W}).
In the effective Hamiltonian $H_{\rm eff}$, the diagonal non-Hermitian term $-i\frac{\kappa_{jj}}{2} \sigma_j^+\sigma_j^-$ denotes the local dissipation of the qubit $Q_j$, while the off-diagonal term $-i\frac{\kappa_{jm}}{2}(\sigma_{j}^+\sigma_m^- + \sigma_j^-\sigma_{m}^+)$ presents the dissipative coupling between the qubits $Q_j$ and $Q_m$~\cite{Harder18,Wang22}. By solving $H_{\rm eff} |\varphi\rangle=E |\varphi\rangle$ in the one-excited subspace, we can obtain the $N$ eigenvectors $\{|\varphi_1\rangle,|\varphi_2\rangle,\cdots,|\varphi_N\rangle\}$ of $H_{\rm eff}$,
\begin{eqnarray}\label{eigenvector}
|\varphi_n\rangle&=&c_{n+1}|\phi_1\rangle-c_1|\phi_{n+1}\rangle,~~~~~n=1,2,\cdots,N-1,\nonumber\\
|\varphi_N\rangle&=&\sum_{j=1}^N c_j|\phi_j\rangle,
\end{eqnarray}
with $|\phi_j\rangle=|0_10_2\cdots 0_{j-1}1_{j}0_{j+1}\cdots 0_N\rangle$, and the corresponding eigenvalues $E_1=E_2=\cdots=E_{N-1}=\omega_q$ and $E_N=\omega_q-i\sum_{j=1}^N \kappa_{jj}/2$. Note that the higher-excited subspace is barely involved in preparing the Bell and $N$-partite $W$ states. Thus, the numerical results related to generating the Bell and $W$ states, which are obtained by numerically solving the master equation (\ref{master-1}) in the whole Hilbert space, can be well explained by using the eigenvectors (\ref{eigenvector}) in the one-excited subspace (cf.~Secs.~\ref{Generation-Bell} and \ref{Generation-W}). This demonstrates the reasonability of the assumption of the one-excited subspace. From ${\rm Im}[E_n]=0$ ($n=1,2,\cdots,N-1$) and ${\rm Im}[E_N] \neq 0$, the decay rate of the eigenvector $|\varphi_n\rangle$ is zero while the decay rate of the eigenvector $|\varphi_N\rangle$ is nonzero. This indicates that the eigenvectors  $\{|\varphi_1\rangle,|\varphi_2\rangle,\cdots,|\varphi_{N-1}\rangle\}$ are dark states of the waveguide QED system, and $|\varphi_N\rangle$ is a bright state. The dark state $|\varphi_n\rangle$ is a steady state of the waveguide QED system due to $\partial \rho / \partial t=0$ with $\rho=|\varphi_n\rangle \langle \varphi_n|$~\cite{Dong12,Zhang15,Zanner22}. In contrast, owing to $\partial \rho / \partial t \neq 0$ with $\rho=|\varphi_N\rangle \langle \varphi_N|$), the bright state  $|\varphi_N\rangle$ is a nonequilibrium state.
In the present work, we use dark states to prepare Bell states and $N$-partite $W$ states in SC waveguide QED. In addition to preparing entangled states, the dark states also have other potential applications in quantum information processing, including protecting quantum systems from dissipations~\cite{Dong12}, building gradient memories to store quantum information~\cite{Zhang15}, realizing quantum information protocols in open quantum systems~\cite{Zanner22}, etc.

\begin{figure}
\includegraphics[width=0.48\textwidth]{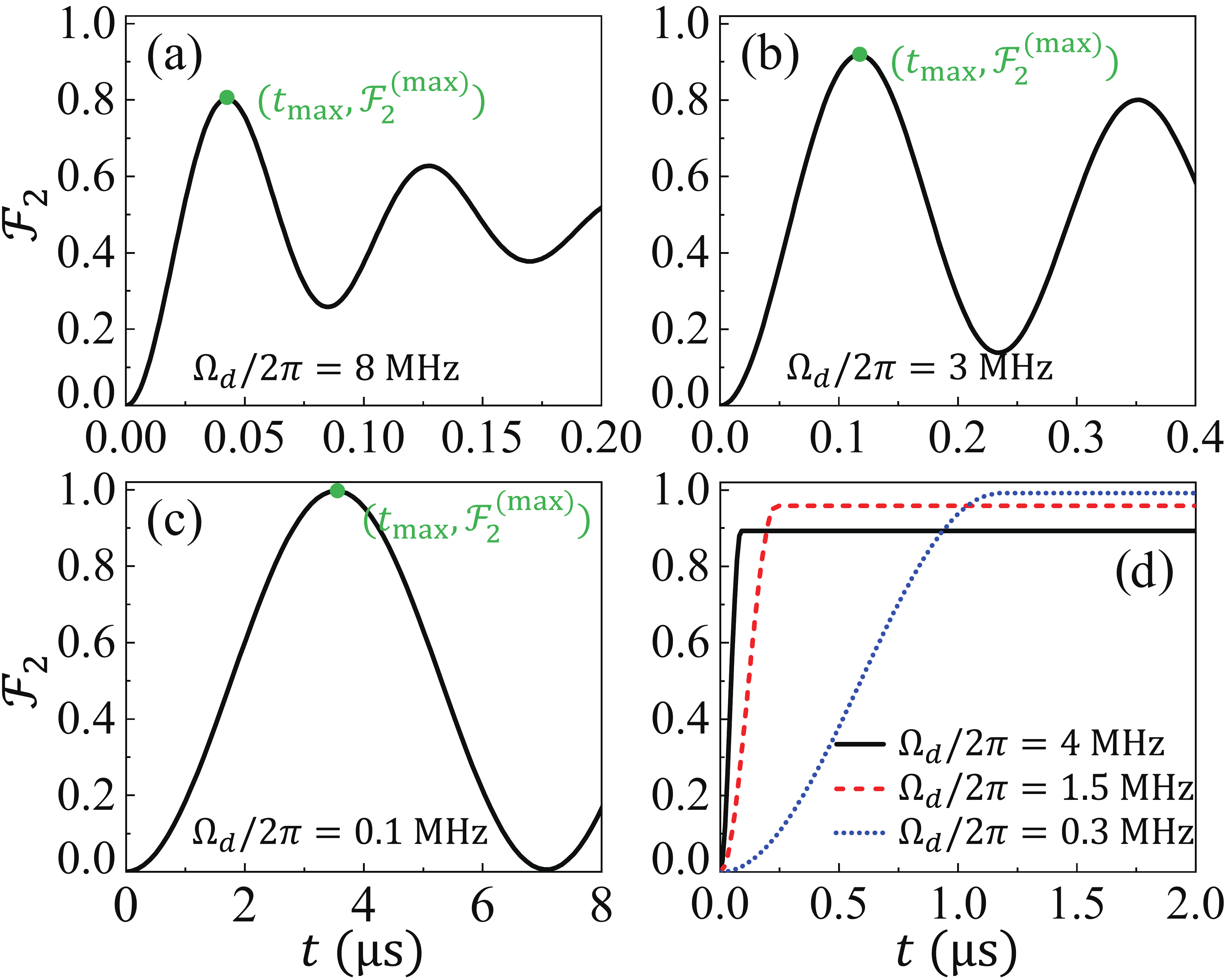}
\caption{Time evolution of the fidelity $\mathcal{F}_2$ of the Bell state $|\Psi_+\rangle$ in the absence of the intrinsic decoherence of the two qubits (i.e., $\gamma=\gamma_\varphi=0$). (a)-(c) Time evolution of the fidelity $\mathcal{F}_2$ for different shapes of the drive pulse, where the duration of the drive pulse is $t_0=+\infty$, while the Rabi frequency is (a) $\Omega_d/2\pi=8$~MHz, (b) $\Omega_d/2\pi=3$~MHz, and (c) $\Omega_d/2\pi=0.1$~MHz. In (a)-(c), the green dots denote the maximum value $\mathcal{F}_2^{\rm (max)}$ of the fidelity $\mathcal{F}_2$ at $t=t_{\rm max}$. (d) Time evolution of the fidelity $\mathcal{F}_2$ for different values of the Rabi frequency $\Omega_d$, where the duration is $t_0=t_{\rm max}$. Here $\Omega_d/2\pi=4$~MHz and $t_0=0.087~\mu$s for the (black) solid curve, $\Omega_d/2\pi=1.5$~MHz and $t_0=0.235~\mu$s for the (red) dashed curve, and $\Omega_d/2\pi=0.3$~MHz and $t_0=1.176~\mu$s for the (blue) dotted curve. Other parameters are chosen as $\omega_q/2\pi=\omega_d/2\pi=5$~GHz, $\kappa_{11}/2\pi=40$~MHz, and $c_1=-c_2=1$.}
\label{fig2}
\end{figure}

\section{Generation of Bell states}\label{Generation-Bell}

In order to generate the Bell states
\begin{eqnarray}\label{Bell-state}
|\Psi_\pm\rangle&=&\frac{1}{\sqrt{2}}(|10\rangle\pm |01\rangle),
\end{eqnarray}
we consider a driven waveguide QED system with two transmon qubits, as schematically shown in Fig.~\ref{fig1}(a). In the one-excited subspace, the considered waveguide QED system has two orthonormal eigenvectors, the dark state $|D\rangle$ and the bright state $|B\rangle$ [cf.~Eq.~(\ref{eigenvector})],
\begin{eqnarray}\label{eigenvector-2}
|D\rangle &=& \frac{1}{\sqrt{c_1^2+c_2^2}\,(c_2/|c_2|)}\left(c_2|10\rangle-c_1|01\rangle\right),\nonumber\\
|B\rangle &=&\frac{1}{\sqrt{c_1^2+c_2^2}\,(c_1/|c_1|)}\left(c_1|10\rangle+c_2|01\rangle\right).
\end{eqnarray}
For $c_1=-c_2$ ($c_1=c_2$), $|D\rangle=|\Psi_+\rangle$ and $|B\rangle=|\Psi_-\rangle$ ($|D\rangle=|\Psi_-\rangle$ and $|B\rangle=|\Psi_+\rangle$).

\begin{figure}
\includegraphics[width=0.48\textwidth]{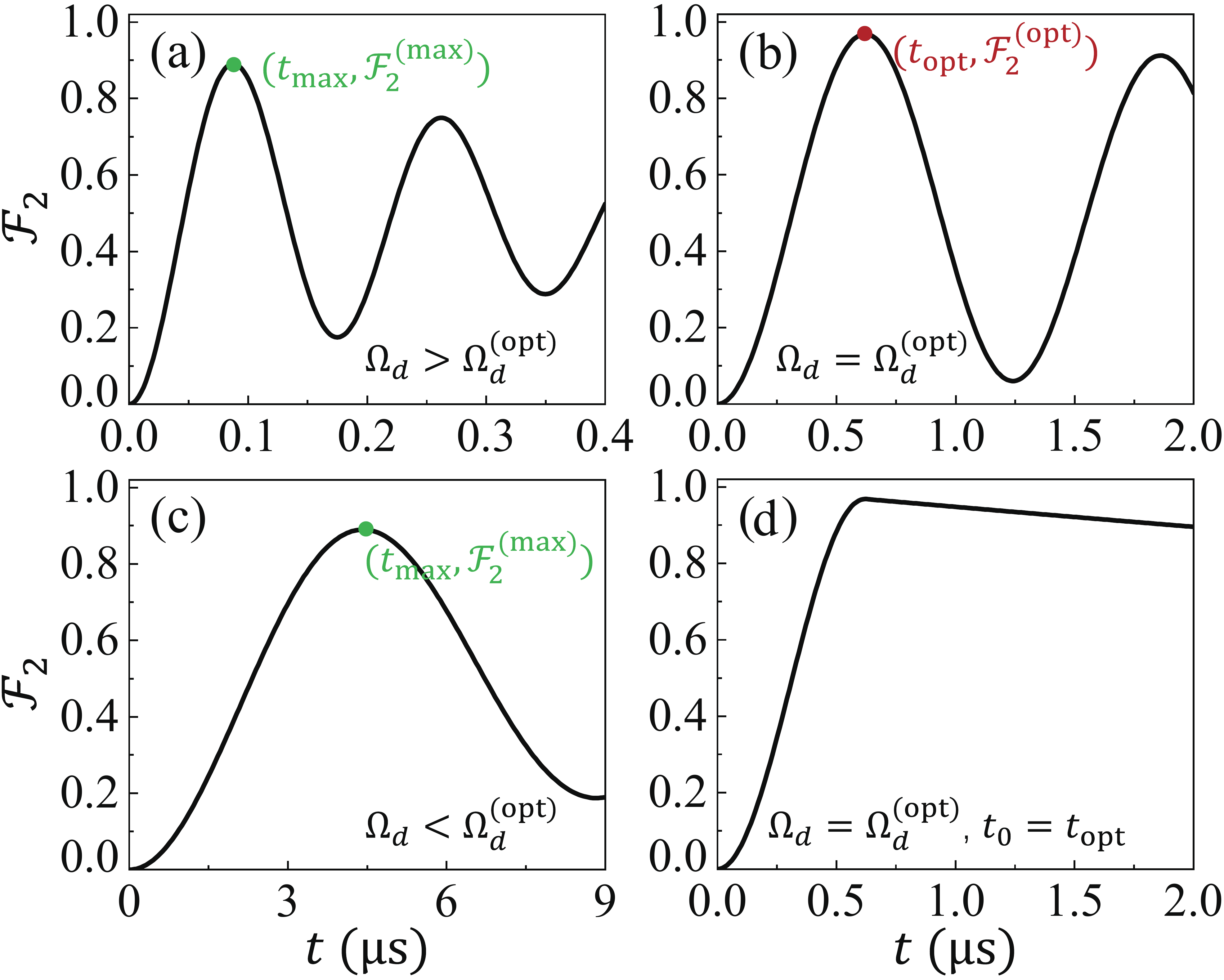}
\caption{Time evolution of the fidelity $\mathcal{F}_2$ of the Bell state $|\Psi_+\rangle$ in the presence of the intrinsic decoherence of the two qubits (e.g., $\gamma^{-1}=60~\mu$s and $\gamma_\varphi^{-1}=25~\mu$s). (a)-(c) Time evolution of the fidelity $\mathcal{F}_2$ for different shapes of the drive pulse, where the duration of the drive pulse is $t_0=+\infty$, while the Rabi frequency is (a) $\Omega_d=7.02\Omega_d^{\rm (opt)}$, (b) $\Omega_d=\Omega_d^{\rm (opt)}$, and (c) $\Omega_d=0.14\Omega_d^{\rm (opt)}$, with $\Omega_d^{\rm (opt)}/2\pi=0.57$~MHz and $t_{\rm opt}=0.623~\mu$s. In (a) and (c), the green dots denote the maximum value $\mathcal{F}_2^{\rm (max)}$ of the fidelity $\mathcal{F}_2$ at $t=t_{\rm max}$, and the red dot denotes the optimal fidelity $\mathcal{F}_2^{\rm (opt)}$ at $t=t_{\rm opt}$ in (b). (d) Time evolution of the fidelity $\mathcal{F}_2$ with $\Omega_d=\Omega_d^{\rm (opt)}$ and $t_0=t_{\rm opt}$. Other parameters are the same as in Fig.~\ref{fig2}.}
\label{fig3}
\end{figure}

For clarity, we first show the numerical results related to preparing the target state $|\Psi_+\rangle$ in the absence of the intrinsic decoherence of the two qubits (i.e., $\gamma=\gamma_\varphi=0$). By numerically solving the master equation in Eq.~(\ref{master-1}) with $c_1=-c_2$~\cite{Johansson12,Johansson13}, we plot the time evolution of the fidelity $\mathcal{F}_2=\rm{Tr}\,\left(\rho|\Psi_+\rangle\langle \Psi_+|\right)$ in Fig.~\ref{fig2}. As shown in Fig.~\ref{fig2}(a), under the drive of the microwave pulse on $Q_1$, the fidelity $\mathcal{F}_2$ evolves from $0$ to its maximum value $\mathcal{F}_2^{\rm (max)}=0.804$ at $t_{\rm max}=0.042~\mu$s (indicated by the green dot), where we assume that the duration of the microwave pulse is infinite, i.e., $t_0=+\infty$. This evolution process can be explained using the eigenvectors of the waveguide QED system in Eq.~(\ref{eigenvector-2}). At the initial time $t=0$, the system of two qubits is in the ground state $|00\rangle$ (corresponding to $\mathcal{F}_2=0$). Due to the drive pulse on $Q_1$, the qubit $Q_1$ will be excited, and the two-qubit system undergoes the state transformation from the ground state $|00\rangle$ to the one-excited state $|10\rangle$. With the eigenvectors $\{|D\rangle,|B\rangle\}$ (with $|D\rangle=|\Psi_+\rangle$ and $|B\rangle=|\Psi_-\rangle$), the state $|10\rangle$ can be expressed as
\begin{eqnarray}\label{eq-10}
|10\rangle=\frac{1}{\sqrt{2}}|D\rangle+\frac{1}{\sqrt{2}}|B\rangle,
\end{eqnarray}
where the component of the bright state $|B\rangle$ will decay to the ground state $|00\rangle$, while the component of the dark state $|D\rangle$ is steady, i.e., the state $|10\rangle$ will evolve to the mixed state of $|00\rangle$ and $|D\rangle$. For the component of $|00\rangle$ in the mixed state, it will be pumped to the one-excited state $|10\rangle$ again, and the process $|00\rangle\rightarrow |10\rangle\rightarrow \{|00\rangle,|D\rangle\}$ is repeated continuously. Ultimately, the waveguide QED system will evolve from the ground state $|00\rangle$ to the dark state $|D\rangle$ (i.e., the Bell state $|\Psi_+\rangle$). It should be noted that the eigenvectors $\{|D\rangle,|B\rangle\}$ are obtained by neglecting the drive pulse. Obviously, if the drive pulse is considered, $|D\rangle$ is not the dark state of the waveguide QED system. As a result,
the maximum fidelity is significantly smaller than one (i.e., $\mathcal{F}_2^{\rm (max)}=0.804 < 1$ ) and $\mathcal{F}_2$ oscillates versus time $t$ [cf.~Fig.~\ref{fig2}(a)]. To decrease this adverse effect, we can use a weaker drive pulse to achieve a higher fidelity $\mathcal{F}_2^{\rm (max)}$, while the time $t_{\rm max}$ for reaching $\mathcal{F}_2^{\rm (max)}$ is longer [see Figs.~\ref{fig2}(b) and \ref{fig2}(c)]. In the weak-drive limit $\Omega_d\rightarrow 0$ with $t_{\rm max}\rightarrow +\infty$, we can in principle generate the Bell state $|\Psi_+\rangle$ with $\mathcal{F}_2^{\rm (max)}=1$. For example, $\mathcal{F}_2^{\rm (max)}=0.997$ for $\Omega_d/2\pi=0.1$~MHz with $t_{\rm max}=3.518~\mu$s. Further, by setting $t_0=t_{\rm max}$, i.e., the duration of the drive pulse is $t_{\rm max}$ rather than infinite, the steady Bell state with a fidelity $\mathcal{F}_2^{\rm (max)}$ can be prepared [see Fig.~\ref{fig2}(d)].

\begin{figure}
\includegraphics[width=0.48\textwidth]{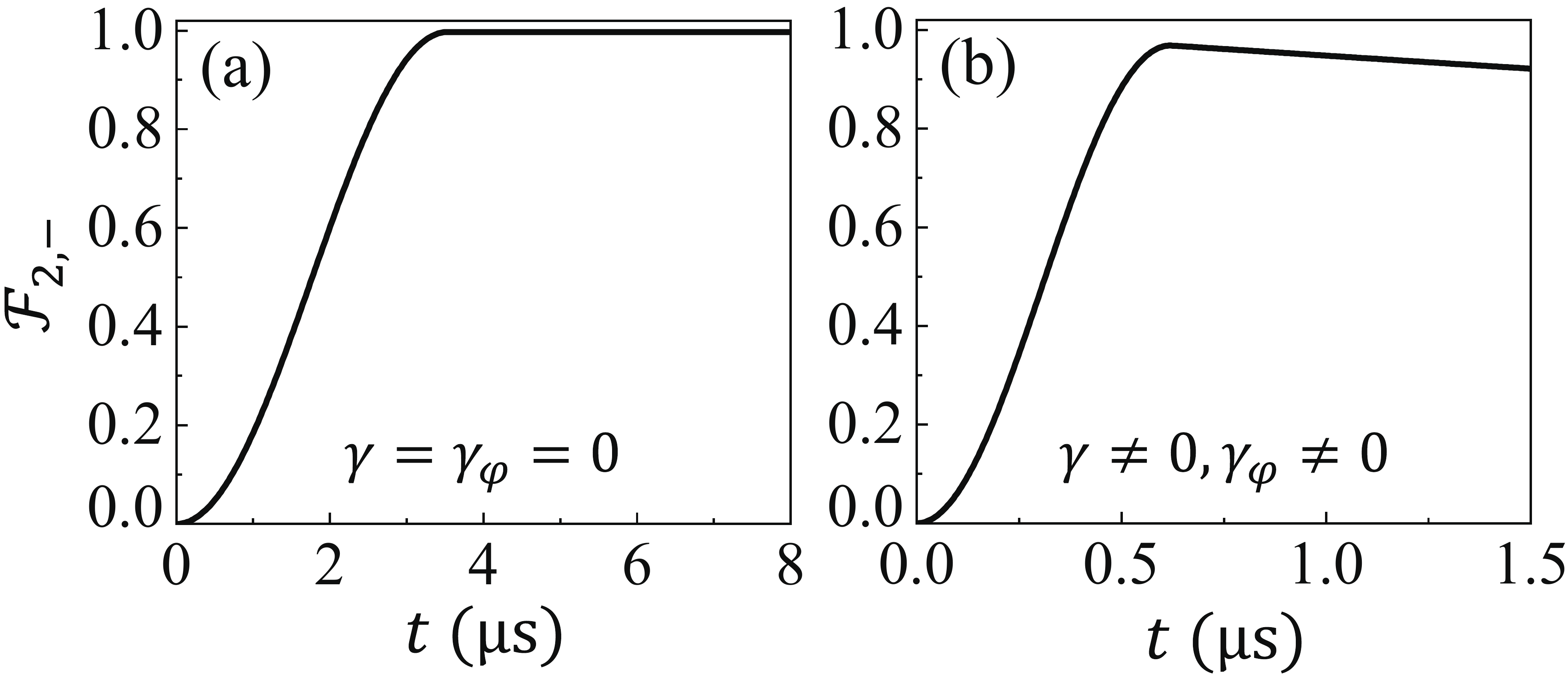}
\caption{Time evolution of the fidelity $\mathcal{F}_{2,-}$ of the Bell state $|\Psi_-\rangle$ in (a) the absence of the intrinsic decoherence of $Q_1$ and $Q_2$ with $\gamma=\gamma_\varphi=0$ and (b) the presence of the intrinsic decoherence of $Q_1$ and $Q_2$ with $\gamma^{-1}=60~\mu$s and $\gamma_\varphi^{-1}=25~\mu$s, where $c_1=c_2=1$. Here $\Omega_d/2\pi=0.1$~MHz and $t_0=3.53~\mu$s in (a), while $\Omega_d/2\pi=0.57$~MHz and $t_0=0.623~\mu$s in (b). Other parameters are the same as in Fig.~\ref{fig2}.}
\label{fig4}
\end{figure}

Further, we consider the intrinsic decoherence of the two qubits, i.e., $\gamma \neq 0$ and $\gamma_\varphi\neq 0$. In this case, there is a specific drive strength $\Omega_d=\Omega_d^{\rm (opt)}$ (with $\Omega_d^{\rm (opt)}/2\pi=0.57$~MHz) to reach the optimal fidelity $\mathcal{F}_2=\mathcal{F}_2^{\rm (opt)}$ (with $\mathcal{F}_2^{\rm (opt)}=0.968$) [cf.~Figs.~\ref{fig3}(a)-\ref{fig3}(c)]. The reason is that if $\Omega_d<\Omega_d^{\rm (opt)}$, the decoherence of $Q_1$ and $Q_2$ will dominate the dynamics of the waveguide QED system and spoil the process of generating the target state $|\Psi_+\rangle$. To obtain a long-lived Bell state $|\Psi_+\rangle$, we set the duration of the drive pulse as $t_0=t_{\rm opt}$. Now the lifetime of the long-lived Bell state $|\Psi_+\rangle$ is determined by the decoherence times of $Q_1$ and $Q_2$ [see Fig.~\ref{fig3}(d)]. Obviously, the lifetime of the long-lived Bell state $|\Psi_+\rangle$ is far longer than that of the transient $|\Psi_+\rangle$ [cf.~Figs.~\ref{fig3}(b) and \ref{fig3}(d)].

With the similar procedures of preparing the Bell state $|\Psi_+\rangle$, we can also generate the other Bell state $|\Psi_-\rangle$ given in Eq.~(\ref{Bell-state}) by setting $c_1=c_2$. In the ideal case without decoherence of $Q_1$ and $Q_2$, the waveguide QED system will evolve from the ground state $|00\rangle$ to the Bell state $|\Psi_-\rangle$ with $\mathcal{F}_{2,-}\rightarrow 1$ in the weak-drive limit $\Omega_d \rightarrow 0$, where $\mathcal{F}_{2,-}=\rm{Tr}\,\left(\rho|\Psi_-\rangle\langle \Psi_-|\right)$ [see Fig.~\ref{fig4}(a)]. Here the generated $|\Psi_-\rangle$ is steady, because it is a dark state of the waveguide QED system. When the decoherence of $Q_1$ and $Q_2$ is considered, the long-lived target state $|\Psi_-\rangle$ with $\mathcal{F}_{2,-}=0.968$ is obtained, where the qubit $Q_1$ is driven by a microwave pulse with an appropriate shape [see Fig.~\ref{fig4}(b)]. Now the lifetime of the generated $|\Psi_-\rangle$ depends on the decoherence times of the two qubits.

\section{Generation of $N$-partite $W$ states}\label{Generation-W}

The scheme proposed in Sec.~\ref{Generation-Bell} can be easily extended to the multi-qubit case for preparing the following $N$-partite $W$ state
\begin{equation}\label{WN-state}
|W_N\rangle=\frac{1}{\sqrt{N}}\sum_{j=1}^N |\phi_j\rangle,~~~N\geq 3.
\end{equation}
Here, we consider $N$ identical transmon qubits coupled to a common waveguide, as depicted in Fig.~\ref{fig1}(b). In the absence of both the drive field and the intrinsic decoherence of qubits, the effective non-Hermitian Hamiltonian $H_{\rm eff}$ of the waveguide QED system in Eq.~(\ref{Hamiltonian-eff}) has $N$ eigenvectors, i.e., the eigenvectors $\{|\varphi_1\rangle,|\varphi_2\rangle,\cdots,|\varphi_{N-1}\rangle\}$ and the eigenvector $|\varphi_N\rangle$ given in Eq.~(\ref{eigenvector}), which are the dark states and the bright state of the waveguide QED system, respectively. While these dark states and the bright state are orthogonal (i.e., $\langle\varphi_n|\varphi_N\rangle=0$), any two of these dark states are not orthogonal (i.e., $\langle\varphi_n|\varphi_m\rangle \neq 0$ with $n \neq m$). Using the Gram-Schmidt orthogonalization, we can obtain a set of orthonormal eigenvectors, $\{|D_1\rangle,\,|D_2\rangle,\,\cdots,\,|D_{N-1}\rangle,\,|B\rangle\}$, where the full expressions for the dark states $\{|D_1\rangle,\,|D_2\rangle,\,\cdots,\,|D_{N-1}\rangle\}$ are cumbersome and not shown, and the bright state $|B\rangle$ is given by
\begin{equation}\label{}
|B\rangle=\frac{1}{\sqrt{\sum_{j=1}^N c_j^2}}\sum_{j=1}^N c_j|\phi_j\rangle.
\end{equation}

\begin{figure}
\includegraphics[width=0.48\textwidth]{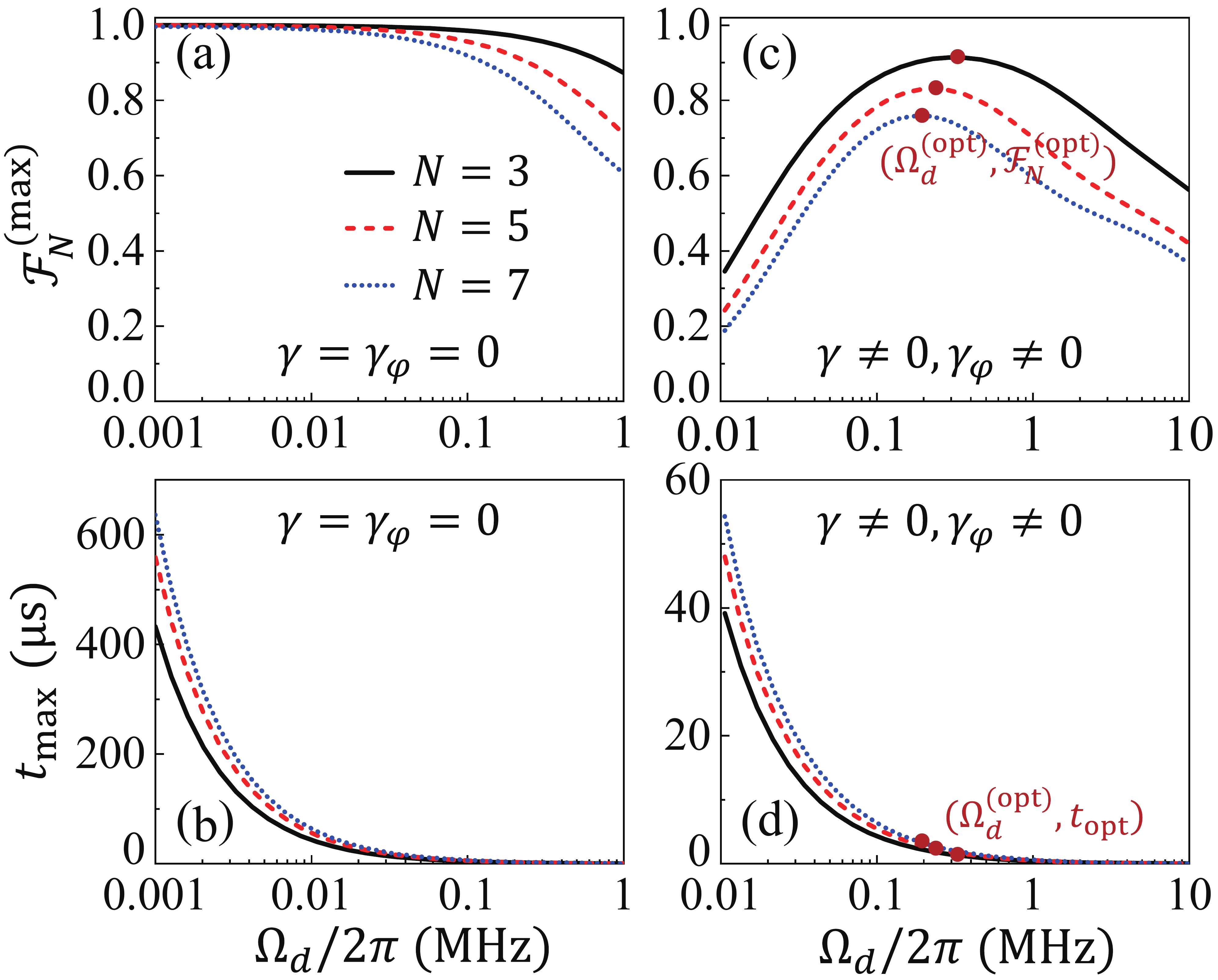}
\caption{The maximum fidelity $\mathcal{F}_{N}^{\rm (max)}$ of the $N$-partite $W$ state $|W_N\rangle$ ($N=3,5,7$) versus the Rabi frequency $\Omega_d/2\pi$ in (a) the absence of the intrinsic decoherence of $N$ qubits with $\gamma=\gamma_\varphi=0$ and (c) the presence of the intrinsic decoherence of $N$ qubits with  $\gamma^{-1}=60~\mu$s and $\gamma_\varphi^{-1}=25~\mu$s, where $c_1=N-1$ and $c_2=c_3=\cdots =c_N = -1$.
For the both cases, the corresponding time $t_{\rm max}$ for reaching $\mathcal{F}_{N}^{\rm (max)}$ is shown in (b) and (d), respectively. In (c), the red dots denote that there exists a specific drive strength $\Omega_d=\Omega_d^{\rm (opt)}$ to realize the optimal fidelity $\mathcal{ F}_N^{\rm (opt)}$, and the needed time $t_{\rm opt}$ for reaching $\mathcal{ F}_N^{\rm (opt)}$ is also marked by the red dots in (d). Other parameters are the same as in Fig.~\ref{fig2}.}
\label{fig5}
\end{figure}

\begin{figure}
\includegraphics[width=0.48\textwidth]{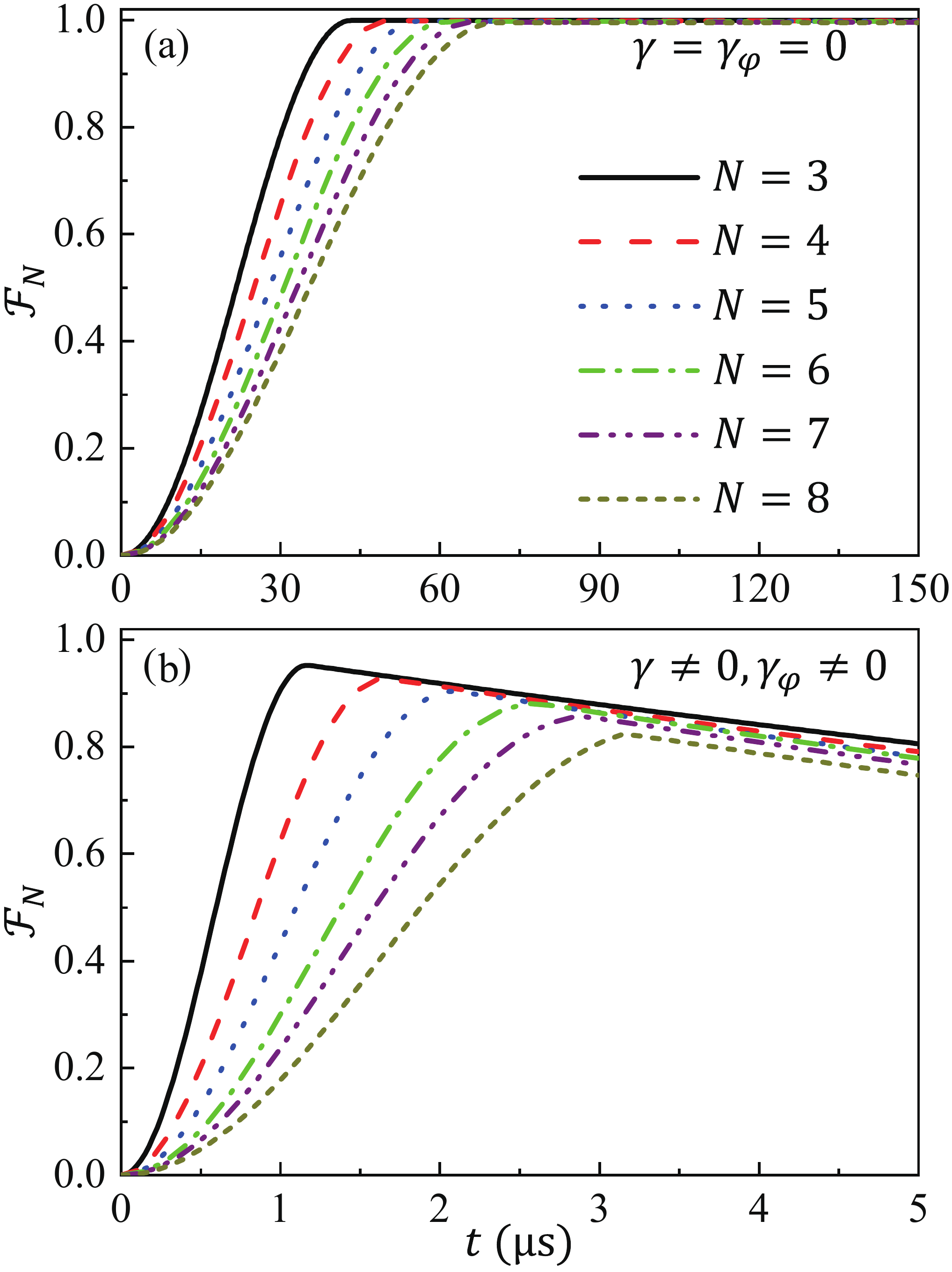}
\caption{Time evolution of the fidelity $\mathcal{F}_N$ of the $N$-partite $W$ state $|W_N\rangle$ ($N=3,4,\cdots,8$) in (a) the absence of the intrinsic decoherence of $N$ qubits with $\gamma=\gamma_\varphi=0$ and (b) the presence of the intrinsic decoherence of $N$ qubits with $\gamma^{-1}=90~\mu$s and $\gamma_\varphi^{-1}=40~\mu$s, where $\kappa_{11}/2\pi=90$~MHz. Here $\Omega_d/2\pi=0.01$~MHz and $t_0=t_{\rm max}$ in (a), while $\Omega_d=\Omega_d^{\rm (opt)}$ and $t_0=t_{\rm opt}$ in (b). The values of $t_{\rm max}$, $\Omega_d^{\rm (opt)}$ and $t_{\rm opt}$ are given in Appendix~\ref{Appendix-B}. Other parameters are the same as in Fig.~\ref{fig5}.}
\label{fig6}
\end{figure}

Now we consider that the $N$ qubits are in the ground state $|0_10_2\cdots 0_N\rangle$ at the initial time $t=0$. By driving the qubit $Q_1$, this qubit is excited and the state of the $N$ qubits becomes the one-excited state $|\phi_1\rangle$, which can be expressed as
\begin{equation}\label{phi1}
|\phi_1\rangle=\sum_{j=1}^{N-1}p_j|D_j\rangle+p_b|B\rangle,
\end{equation}
with $p_j=\langle D_j |\phi_1\rangle$ and  $p_b=\langle B |\phi_1\rangle=c_1/\sqrt{\sum_{j=1}^N c_j^2}$. For convenience, we define a dark state $|\mathcal{D}\rangle\equiv\sum_{j=1}^{N-1}p_j|D_j\rangle=|\phi_1\rangle-p_b|B\rangle$. In the basis of $|\phi_j\rangle$, the dark state $|\mathcal{D}\rangle$ can be rewritten as
\begin{equation}\label{dark-state}
\begin{split}
|\mathcal{D}\rangle=\sum_{j=1}^N f_j|\phi_j\rangle,
\end{split}
\end{equation}
where
\begin{eqnarray}\label{}
f_1&=&1-\frac{c_1^2}{\sum_{j=1}^N c_j^2},\nonumber\\
f_j&=&-\frac{c_1c_j}{\sum_{j=1}^N c_j^2},~~~~~~j=2,3,\cdots,N.
\end{eqnarray}
If $f_1=f_2=\cdots=f_{N}$, the dark state $|\mathcal{D}\rangle$ can be related to a standard $N$-partite $W$ state $|W_N\rangle$ given in Eq.~(\ref{WN-state}), i.e., $|\mathcal{D}\rangle \propto |W_N\rangle$. Solving $f_1=f_2=\cdots=f_{N}$, we have
\begin{eqnarray}\label{c1-cn}
c_1&=&-(N-1)c_2,\nonumber\\
c_2&=&c_3=\cdots=c_{N},
\end{eqnarray}
corresponding to $f_1=f_2=\cdots =f_{N}=1/N$. When $c_j$ satisfies Eq.~(\ref{c1-cn}), the dark state $|\mathcal{D}\rangle$ in Eq.~(\ref{dark-state}) is reduced to $|\mathcal{D}\rangle=\left(1/\sqrt{N}\right)|W_N\rangle$. Now Eq.~(\ref{phi1}) can be rewritten as
\begin{equation}\label{}
|\phi_1\rangle=\frac{1}{\sqrt{N}}|W_N\rangle-\frac{c_2}{|c_2|}\sqrt{1-\frac{1}{N}}|B\rangle,
\end{equation}
which has the same form as Eq.~(\ref{eq-10}). Here the component of the dark state $|W_N\rangle$ in $|\phi_1\rangle$ is steady, but the component of the bright state $|B\rangle$ will decay to the ground state $|0_10_2\cdots 0_N\rangle$. Therefore, we can generate the $N$-partite $W$ state $|W_N\rangle$ by setting $c_1=-(N-1)c_2$ with $c_2=c_3=\cdots=c_{N}$ and driving the qubit $Q_1$ with an appropriate microwave pulse.

To demonstrate the above analyses, we take the cases of $N=3,$ $5,$ and $7$ as an example. We plot Fig.~\ref{fig5}, which shows the maximum value $\mathcal{F}_{N}^{\rm (max)}$ of the fidelity $\mathcal{F}_N=\rm{Tr}\,\left(\rho|W_N\rangle\langle W_N|\right)$ of the $N$-partite $W$ state $|W_N\rangle$ and the time $t_{\rm max}$ for reaching $\mathcal{F}_{N}^{\rm (max)}$ as functions of the Rabi frequency $\Omega_d/2\pi$. It should be emphasized that the following discussions and conclusions related to Fig.~\ref{fig5} are valid for any value of $N$. For $N=3,$ $5,$ and $7$, both the maximum fidelity $\mathcal{F}_{N}^{\rm (max)}$ and the corresponding time $t_{\rm max}$ versus $\Omega_d/2\pi$ monotonically decrease in the ideal case without the intrinsic decoherence of qubits [see Figs.~\ref{fig5}(a) and \ref{fig5}(b)]. In the weak-drive limit $\Omega_d/\kappa_{jj}\rightarrow 0$, one has $\mathcal{F}_{N}^{\rm (max)}\rightarrow 1$ but $t_{\rm max}\rightarrow +\infty$. By choosing the duration $t_0=t_{\rm max}$ of the weak drive pulse (with, e.g., $\Omega_d/2\pi=0.01$~MHz), a steady $W$ state $|W_N\rangle$ with fidelity $\mathcal{F}_{N}=\mathcal{F}_{N}^{\rm (max)}$ ($\geq 0.995$) can be prepared [cf.~Fig.~\ref{fig6}(a)].

When the intrinsic decoherence of qubits is considered, there is a specific value $\Omega_d=\Omega_{d}^{\rm (opt)}$ of the Rabi frequency for obtaining the optimal fidelity $\mathcal{F}_{N}=\mathcal{F}_{N}^{\rm (opt)}$ at time $t=t_{\rm opt}$ [see Figs.~\ref{fig5}(c) and \ref{fig5}(d)]. This characteristic is similar to the case of $N=2$ [cf.~Figs.~\ref{fig3}(a)-\ref{fig3}(c)]. Under the action of the drive pulse with $\Omega_d=\Omega_{d}^{\rm (opt)}$ and $t_0=t_{\rm opt}$ on $Q_1$, the system of $N$ qubits evolves from the ground state $|0_10_2\cdots 0_N\rangle$ to the target state $|W_N\rangle$ with fidelity $\mathcal{F}_{N}=\mathcal{F}_{N}^{\rm (opt)}$ [cf.~Fig.~\ref{fig6}(b)]. Contrary to the ideal case of $\gamma=\gamma_\varphi=0$, the intrinsic decoherence of qubits limits the lifetime of the generated $W$ state $|W_N\rangle$, and the optimal fidelity $\mathcal{F}_{N}^{\rm (opt)}$ is smaller than the ideal value 1.
Interestingly, we find that after the drive pulse are shut down, the fidelity $\mathcal{F}_N$ versus time decrease at the same rate for different qubit number $N$. This phenomenon can be explained using an effective non-Hermitian Hamiltonian. Because the generated $W$ state $|W_N\rangle$ is a dark state that decouples from the waveguide, here we only consider the intrinsic decoherence of qubits. Now we can write the effective non-Hermitian Hamiltonian of $N$ qubits as $\mathcal{H}_{\rm eff}=\sum_{j=1}^{N}(\omega_q-i\gamma/2)\sigma_j^+\sigma_j^- - i\gamma_\varphi\sigma_j^z\sigma_j^z/4$ ~\cite{Minganti19}. Under this effective Hamiltonian, the system of $N$ qubits evolves from the initial state $|\psi(0)\rangle=|W_N\rangle$ to $|\psi(t)\rangle=e^{-i\mathcal{H}_{\rm eff}t}|\psi(0)\rangle=e^{-i\omega_qt-(\gamma/2+\gamma_\varphi/4)t}|W_N\rangle$ at time $t$. Correspondingly, the time evolution of the fidelity is given by $\mathcal{F}_N=|\langle W_N|\psi(t)\rangle|^2=e^{-(\gamma+\gamma_\varphi/2)t}$, which is independent of the qubit number $N$.

\begin{figure}
\includegraphics[width=0.48\textwidth]{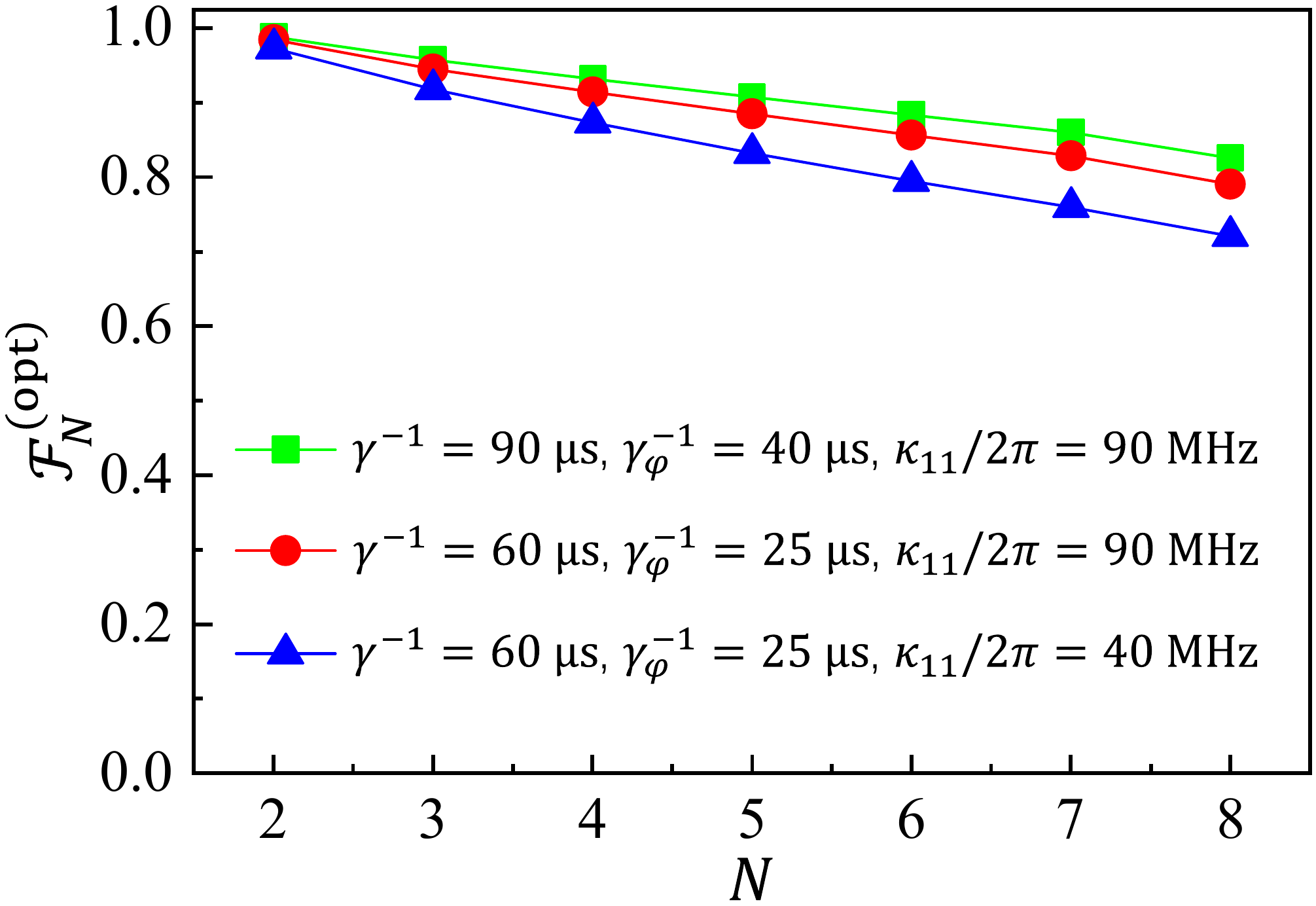}
\caption{The optimal fidelity $\mathcal{F}_{N}^{\rm (opt)}$ of the Bell state $|\Psi_+\rangle$ and $N$-partite $W$ states $|W_N\rangle$ ($N=3,4,\cdots,8$) versus the qubit number $N$ for different energy relaxation rate $\gamma$, pure dephasing rate $\gamma_\varphi$ and local dissipation rate $\kappa_{11}$, where $\Omega_d=\Omega_d^{\rm (opt)}$ and $t_0=t_{\rm opt}$. The values of $\Omega_d^{\rm (opt)}$ and $t_{\rm opt}$ can be found in Appendix~\ref{Appendix-B}. Other parameters are the same as in Fig.~\ref{fig5}.}
\label{fig7}
\end{figure}

\section{Discussions and conclusions}\label{summary}

It should be noted that the optimal fidelity $\mathcal{F}_{N}^{\rm (opt)}$ of the Bell and $N$-partite $W$ states versus $N$ monotonically decreases
when the intrinsic decoherence of qubits is considered [cf.~Figs.~\ref{fig6}(b) and \ref{fig7}]. This is because the used value of $\kappa_{11}$ is the same, while the parameter $\kappa_{jj}=\kappa_{11}/(N-1)^2$ ($j\geq 2$) depends on the qubit number $N$ (satisfying $\kappa_{jj}\leq\kappa_{11}$). For given values of $\kappa_{11}$, $\gamma$ and $\gamma_\varphi$, a smaller $N$ (corresponding to a larger $\kappa_{jj}$) makes the conditions $\gamma/\kappa_{jj}\rightarrow 0$ and $\gamma_\varphi/\kappa_{jj}\rightarrow 0$ better satisfied, which results in a larger $\mathcal{F}_{N}^{\rm (opt)}$. In the limit $\gamma/\kappa_{jj}=\gamma_\varphi/\kappa_{jj} = 0$, $\mathcal{F}_{N}^{\rm (opt)}\approx 1$ can be obtained [cf.~Fig.~\ref{fig6}(a)]. Therefore, as shown in Fig.~\ref{fig7}, the optimal fidelity $\mathcal{F}_{N}^{\rm (opt)}$ for a specific $N$ can be improved by increasing the coupling strength of the SC qubit to the waveguide (i.e., larger values of $\kappa_{jj}$)~\cite{Hoi11,Diaz17,Mirhosseini19} or using SC qubits with longer decoherence times (i.e., smaller values of $\gamma$ and $\gamma_\varphi$)~\cite{Larsen15,You07,Earnest18,Siddiqi21}.

Considering the experimentally accessible parameters, we choose $\omega_q/2\pi=5$~GHz, $\kappa_{jj}/2\pi \leq 90$~MHz, $\gamma^{-1} \leq 90$~$\mu s$, $\gamma_\varphi^{-1} \leq 40$~$\mu s$, and $\Omega_d/2\pi\leq 10$~MHz in the numerical simulations (cf. Figs.~\ref{fig2}-\ref{fig7}). Note that the specific drive strength $\Omega_d^{\rm (opt)}$ (for reaching the optimal fidelity $\mathcal{F}_N^{\rm (opt)}$) is smaller than 1~MHz, cf. Table~\ref{Table-C2} in Appendix~\ref{Appendix-B}. Experimentally, the strong coupling between the transmon qubit and the waveguide has been demonstrated, where the decay rate $\kappa_{jj}/2\pi=99.5$~MHz of a transmon qubit, solely induced by a waveguide, is reported~\cite{Mirhosseini19}. In addition, the typical frequency of a transmon qubit is on the order of 1-10~GHz, and the decoherence times $\gamma^{-1}$ and $\gamma_\varphi^{-1}$ of the state-of-the-art transmon qubit are on the order of 10-100~$\mu$s~\cite{Gu17,Rigetti12,Chang13,Wei20}. As for the microwave field acting on the qubit, the drive strength of $10$~MHz can be easily obtained in experiments~\cite{Gu17}. With these experimentally accessible conditions, our scheme can be implemented in the SC waveguide QED system.
If transmon qubits are replaced by other types of SC qubits, our scheme is also valid. However, compared with other types of SC qubits, the transmon qubit has its unique advantages, e.g., the simplicity and the flexibility of circuit architectures~\cite{Huang20}. Currently, the transmon qubit is the most popular SC qubit. Thus, we use transmon qubits in our scheme.

In our scheme, the generated Bell states $|\Psi_\pm\rangle$ and $N$-partite $W$ state $|W_N\rangle$ are in the one-excited subspace. On the contrary, other two Bell states $|B_\pm\rangle=\frac{1}{\sqrt{2}}(|11\rangle \pm |00\rangle)$ and $N$-partite GHZ states involve the higher-excited states $|11\rangle$ and $|1_11_2\cdots1_N\rangle$, which are not in the one-excited subspace. The higher-excited states $|11\rangle$ and $|1_11_2\cdots1_N\rangle$ can leak into the one-excited subspace, where there are many stable dark states. This implies that it is difficult to engineer Bell states $|B_\pm\rangle$ and $N$-partite GHZ states using the present scheme.
In addition, it is worth noting that we have ignored the non-Markovian effect in our scheme [cf. Eq.~(\ref{master-1}) and related discussions], which requires that the distance between any two transmon qubits cannot be too long (i.e., $|x_j-x_m|\ll 10$~m). If the distance between two transmon qubits is on the order of $10$~m, the non-Markovian effect in this system should be considered~\cite{Kannan20}. The impact of non-Markovian effect on our scheme is an interesting and important topic, which will be investigated in the future.

To summarize, we have presented a novel scheme for preparing the Bell and $N$-partite $W$ states of long-distance SC qubits in a SC waveguide. With appropriate system parameters, we find that the Bell state or the $N$-partite $W$ state is a dark state of the considered SC waveguide QED system. If we drive one of qubits by a proper microwave pulse, the qubits will evolve from their ground states to the Bell state or the $W$ state. In the ideal case without decoherence of qubits, the prepared Bell and $W$ states are steady since they are decoupled from the waveguide. When the decoherence of SC qubits is considered, the lifetimes of the Bell and $W$ states are on the order of the decoherence times of the qubits. The generated Bell and $N$-partite $W$ states can be used to entangle long-distance nodes in waveguide QED networks, which is important in quantum information processing.

\section*{Acknowledgments}

This work is supported by the National Natural Science Foundation of China (Grants No.~12205069, No.~12204139, and No.~U21A20436), the Key-Area Research and Development Program of GuangDong province (Grant No.~2018B030326001), and the key program of the Natural Science Foundation of Anhui (Grant No.~KJ2021A1301).

\appendix

\section{The master equation of the waveguide QED system with transmon qubits}\label{Appendix-A}

As shown in Fig.~\ref{fig1}, $N$ transmon qubits are coupled to a 1D waveguide (i.e., a SC microwave transmission line) with a continuum of left- and right-propagating microwave modes. For the coupled system, the total Hamiltonian $H_{\rm tot}$ contains three parts:
\begin{eqnarray}\label{}
H_{\rm tot}&=&H_{\rm sys}+H_{\rm res}+H_{\rm int}.
\end{eqnarray}
Here, $H_{\rm sys}$ is the Hamiltonian of $N$ transmon qubits given by
\begin{eqnarray}\label{}
H_{\rm sys}&=&\sum_{j=1}^N \omega_j \sigma_j^+\sigma_j^-,
\end{eqnarray}
where $\omega_j$ is the transition frequency between the ground state $|0_j\rangle$ and the excited state $|1_j\rangle$ of $Q_j$, and $\sigma_j^-=|0_j\rangle\langle 1_j|$ and $\sigma_j^+=|1_j\rangle\langle 0_j|$ are the lowering and raising operators of $Q_j$, respectively. The Hamiltonian $H_{\rm res}$ of the waveguide reads~\cite{Shen05}
\begin{eqnarray}\label{}
H_{\rm res}&=&\int^{+\infty}_0 d\nu_k\,\nu_k[b^\dag_L(\nu_k)b_L(\nu_k)+b^\dag_R(\nu_k)b_R(\nu_k)],
\end{eqnarray}
where the creation and annihilation operators $b_L(\nu_k)$ and $b^\dag_L(\nu_k)$ [$b_R(\nu_k)$ and $b^\dag_R(\nu_k)$] denote the leftward (rightward) travelling microwave mode at frequency $\nu_k$ in the waveguide. Lastly, the electric-dipole interactions between the $N$ qubits and the continuum of propagating microwave modes in the waveguide can be expressed as~\cite{Lalumiere13}
\begin{eqnarray}\label{}
H_{\rm int}&=&\sum_{j=1}^N\int^{+\infty}_0 d\nu_k\, g_j\sqrt{\nu_k}[b(\nu_k)\sigma^+_j+b^\dag(\nu_k)\sigma^-_j],
\end{eqnarray}
with $b(\nu_k)=b_L(\nu_k)e^{-iv_kx_j/\upsilon}+b_R(\nu_k)e^{iv_kx_j/\upsilon}$, where $g_j$ is the coupling strength of $Q_j$ to the waveguide, $x_j$ is the location of $Q_j$, and $\upsilon$ is the velocity of the microwave in the waveguide.

In the coupled system, the waveguide with continuous propagating modes $b_L(\nu_k)$ and $b_R(\nu_k)$ acts as a common reservoir. At zero temperature, we take trace over the degrees of freedom of the waveguide under both the Born approximation and the Markov approximation. The Born-Markovian master equation for the $N$ qubits can be derived as~\cite{Lalumiere13}
\begin{equation}\label{Eq-A1}
 \frac{\partial \rho}{\partial t} =-i[H_{\rm sys}+H_{\rm qq},\rho]
                                    +\sum_{j=1}^{N}\sum_{m=1}^{N}\frac{\kappa_{jm}}{2}[2\sigma_j^-\rho\sigma_{m}^+ -\sigma_{m}^+\sigma_j^-\rho
                                                 -\rho\sigma_{m}^+\sigma_j^-],
\end{equation}
where $\rho$ is the reduced density operator of the $N$ qubits,
\begin{equation}\label{Eq-A2}
H_{\rm qq}=\sum_{j=1}^N\sum_{m=1}^N\lambda_{jm}\sigma_j^-\sigma_{m}^+
\end{equation}
describes the interactions among $N$ transmon qubits due to the waveguide, and $\lambda_{jm}$ ($\kappa_{jm}$) is the coherent coupling strength between (cooperative decay rate of) $Q_j$ and $Q_m$ given by
\begin{eqnarray}\label{coupling-decay}
\lambda_{jm}&=&-i\pi g_j g_{m}\left(\omega_j e^{ik_j|x_j-x_{m}|}-\omega_{m} e^{-ik_{m}|x_j-x_{m}|}\right),\nonumber\\
\kappa_{jm}&=&2\pi g_jg_{m}\left(\omega_j e^{ik_j|x_j-x_{m}|}+\omega_{m} e^{-ik_{m}|x_j-x_{m}|}\right).
\end{eqnarray}
Here, $k_j=\omega_j/\upsilon$ is the wave vector of the microwave at frequency $\omega_j$. Note that the Markov approximation is reasonable when the distance between any two transmon qubits is on the order of meters~\cite{Kannan20}. From Eq.~(\ref{Eq-A1}), we find that the waveguide has two effects on $N$ qubits. On the one hand, the waveguide mediates the coherent couplings between different qubits and induces the frequency shifts of qubits [corresponding to the Hamiltonian $H_{\rm qq}$ in Eq.~(\ref{Eq-A1})]. On the other hand, the waveguide can cause the cooperative dissipations of different qubits and the local dissipations of qubits [cf. the second term in Eq.~(\ref{Eq-A1})].

\newcommand{\tabincell}[2]{}
\renewcommand\tabcolsep{5.5pt}
\begin{table}[h]
	\centering
	\scriptsize
	\caption{The values of $t_{\rm max}$, $\Omega_d^{\rm (opt)}$ and $t_{\rm opt}$ are used in Fig.~\ref{fig6}.}
	\label{tab:notations}
	\begin{tabular}{ccccccc}
		\\[-1mm]
		\hline
		\hline\\[-1mm]
		 qubit number $N$ &3&4&5&6&7&8\\[1mm]
		\hline
		\vspace{-3mm}\\[2mm]
		$t_{\rm max}$ ($\mu$s)   &43.3&50.1&55.8&61.4&66.1&70.9\\[2mm]
		$\Omega_d^{\rm (opt)}/2\pi$ (MHz)   &0.370&0.301&0.265&0.233&0.221&0.202\\[2mm]
        $t_{\rm opt}$ ($\mu$s)    &1.165&1.645&2.09&2.606&2.878&3.149\\[2mm]
        \hline
		\hline
	\end{tabular}
\label{Table-C1}
\end{table}

When the $N$ qubits are resonant, i.e., $\omega_1=\omega_2=\cdots=\omega_N=\omega_q$, $\lambda_{jm}$ and $\kappa_{jm}$ in Eq.~(\ref{coupling-decay}) become
\begin{eqnarray}\label{}
\lambda_{jm}&=&2\pi g_jg_{m}\omega_q \sin(k|x_j-x_{m}|),\nonumber\\
\kappa_{jm}&=&4\pi g_jg_{m}\omega_q \cos (k|x_j-x_{m}|),
\end{eqnarray}
with $k=k_1=k_2=\cdots=k_N=\omega_q/\upsilon$. Obviously, $\lambda_{jm}=\lambda_{mj}$ and $\kappa_{jm}=\kappa_{mj}$. Further, in the case of $kx_j=\pm l\pi$ ($l=0,1,2,\cdots$), one has $\lambda_{jm}=0$ but $\kappa_{jm}$ can be expressed as
\begin{equation}\label{}
\kappa_{jm}=4\pi g_j g_{m}\omega_q e^{ik(x_j-x_{m})}=2c_jc_{m}\kappa ,\\
\end{equation}
where we have introduced the collective decay rate $\kappa=2\pi g^2\omega_q$ of $N$ qubits, the collective coupling strength $g$ of $N$ qubits to the waveguide, and the relative coupling strength $c_j=(g_j/g) e^{ikx_j}$ of $Q_j$ to the waveguide. Note that $c_j$ is real (i.e., $c_j=c_j^*$) due to $e^{ikx_j}=\pm 1$. Now the coherent coupling between $Q_j$ and $Q_m$ disappears due to $\lambda_{jm}=0$, and only the cooperative dissipation of $Q_j$ and $Q_m$ exists with $\kappa_{jm}\neq 0$. In this case, Eq.~(\ref{Eq-A1}) is reduced to the master equation in Eq.~(\ref{master-1}) without both the drive pulse and the intrinsic decoherence of $N$ qubits (i.e., $\Omega_d=\gamma=\gamma_\varphi=0$).

\section{The parameters used in Figs.~\ref{fig6} and \ref{fig7}}\label{Appendix-B}

In Table~\ref{Table-C1}, we give the values of $t_{\rm max}$, $\Omega_d^{\rm (opt)}$ and $t_{\rm opt}$ used in Fig.~\ref{fig6}. In addition, the values of $\Omega_d^{\rm (opt)}$ and $t_{\rm opt}$ used in Fig.~\ref{fig7} are also given in Table~\ref{Table-C2}.

\begin{table}[h]
	\centering
	\scriptsize
	\caption{The values of $\Omega_d^{\rm (opt)}$ and $t_{\rm opt}$ are used in Fig.~\ref{fig7}, where the second and third lines correspond to the case of $\gamma^{-1}=60~\mu$s, $\gamma^{-1}_\varphi=25~\mu$s and $\kappa_{11}/2\pi=40$~MHz, the fourth and fifth lines correspond to the case of $\gamma^{-1}=60~\mu$s, $\gamma^{-1}_\varphi=25~\mu$s and $\kappa_{11}/2\pi=90$~MHz, and the sixth and seventh lines correspond to the case of  $\gamma^{-1}=90~\mu$s, $\gamma^{-1}_\varphi=40~\mu$s and $\kappa_{11}/2\pi=90$~MHz.}
	\label{tab:notations}
	\begin{tabular}{cccccccc}
		\\[-1mm]
		\hline
		\hline\\[-1mm]
		 qubit number $N$ &2&3&4&5&6&7&8\\[1mm]
		\hline
		\vspace{-3mm}\\[2mm]
		$\Omega_d^{\rm (opt)}/2\pi$ (MHz)   &0.570&0.310&0.257&0.227&0.205&0.188&0.180\\[2mm]
        $t_{\rm opt}$ ($\mu$s)   &0.623 &1.390&1.924&2.431&2.925&3.394&3.530\\[2mm]
        \hline
		\vspace{-3mm}\\[2mm]
		$\Omega_d^{\rm (opt)}/2\pi$ (MHz)   &0.86&0.47&0.37&0.333&0.306&0.279&0.224\\[2mm]
        $t_{\rm opt}$ ($\mu$s)   &0.413 &0.917&1.338&1.662&1.974&2.279&2.842\\[2mm]
        \hline
		\vspace{-3mm}\\[2mm]
		$\Omega_d^{\rm (opt)}/2\pi$ (MHz)   &0.69&0.37&0.301&0.265&0.233&0.221&0.202\\[2mm]
        $t_{\rm opt}$ ($\mu$s)   &0.515 &1.165&1.645&2.09&2.606&2.878&3.149\\[2mm]
        \hline
		\hline
	\end{tabular}
\label{Table-C2}
\end{table}

\section{The effects of the position deviations and the higher levels of transmon qubits on our scheme}\label{Appendix-C}

\begin{figure}
\includegraphics[width=0.48\textwidth]{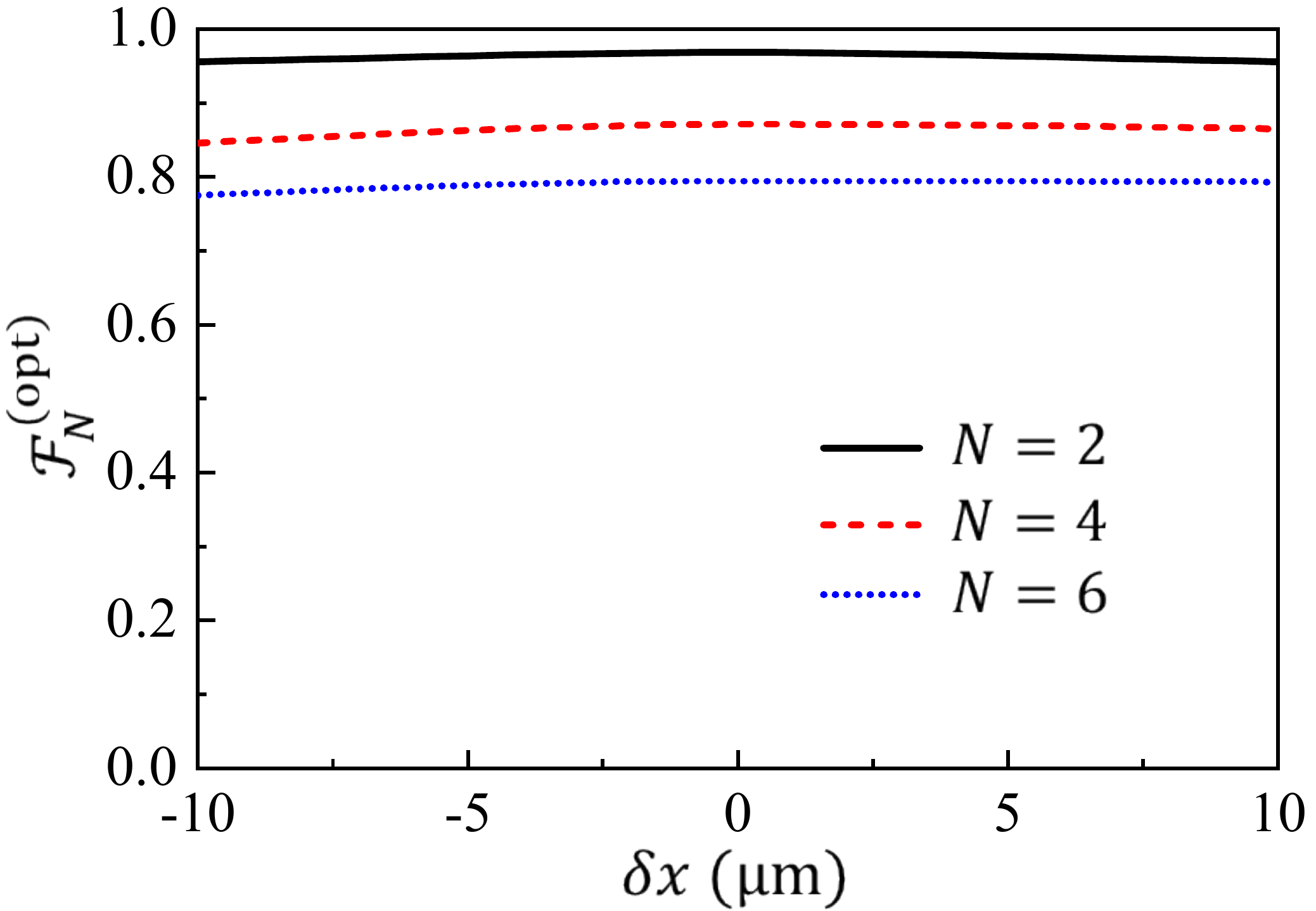}
\caption{The optimal fidelity $\mathcal{F}_{N}^{\rm (opt)}$ of the Bell state $|\Psi_+\rangle$ and $N$-partite $W$ state $|W_N\rangle$ ($N=4,6$) versus the position deviation $\delta x$ of the second qubit $Q_2$, obtained by numerically solving Eq.~(\ref{Eq-A1}). In the numerical simulation, the location of $Q_2$ is set as $x_2=\pi/k+\delta x$, and other parameters are the same as in Fig.~\ref{fig5}(c).}
\label{fig8}
\end{figure}

\subsection{The effect of the position deviations of transmon qubits on our scheme}\label{Appendix-C1}

To prepare Bell states and $N$-partite $W$ states in our scheme, the location of the qubit $Q_j$ is given by $x_j=\pm l\pi/k=\pm l \lambda_0$, where $\lambda_0=\pi/k=\pi\upsilon/\omega_q$. Given that the velocity of the microwave in the waveguide is $\upsilon=10^8$~m/s~\cite{Lalumiere13}, we find $\lambda_0=1$~cm for $\omega_q/2\pi=5$~GHz. With the experimentally accessible technologies, the position deviation $\delta x$ of the SC qubits from the ideal position $x_j=\pm l \lambda_0$ can be controlled within 1~$\mu$m (i.e., $\delta x < 1$~$\mu$m). Since the position deviation $\delta x$ is very small ($\delta x/\lambda_0< 10^{-4}$), its effect on our scheme is tiny.

To demonstrate the above analyses, we consider an example, i.e., the second qubit $Q_2$ deviates from the ideal position as an example. In Fig.~\ref{fig8}, we plot the optimal fidelity $\mathcal{F}_{N}^{\rm (opt)}$ of the Bell state $|\Psi_+\rangle$ and $N$-partite $W$ state $|W_N\rangle$ ($N=4,6$) as a function of the position deviation $\delta x$ of $Q_2$, where the position of $Q_2$ is given by $x_2=\pi/k+\delta x$. It can be seen that $\mathcal{F}_{N}^{\rm (opt)}$ versus $\delta x$ varies slowly around the ideal value $\delta x=0$. This implies that our scheme is robust to small position deviations of the qubits.

\begin{figure}
\includegraphics[width=0.48\textwidth]{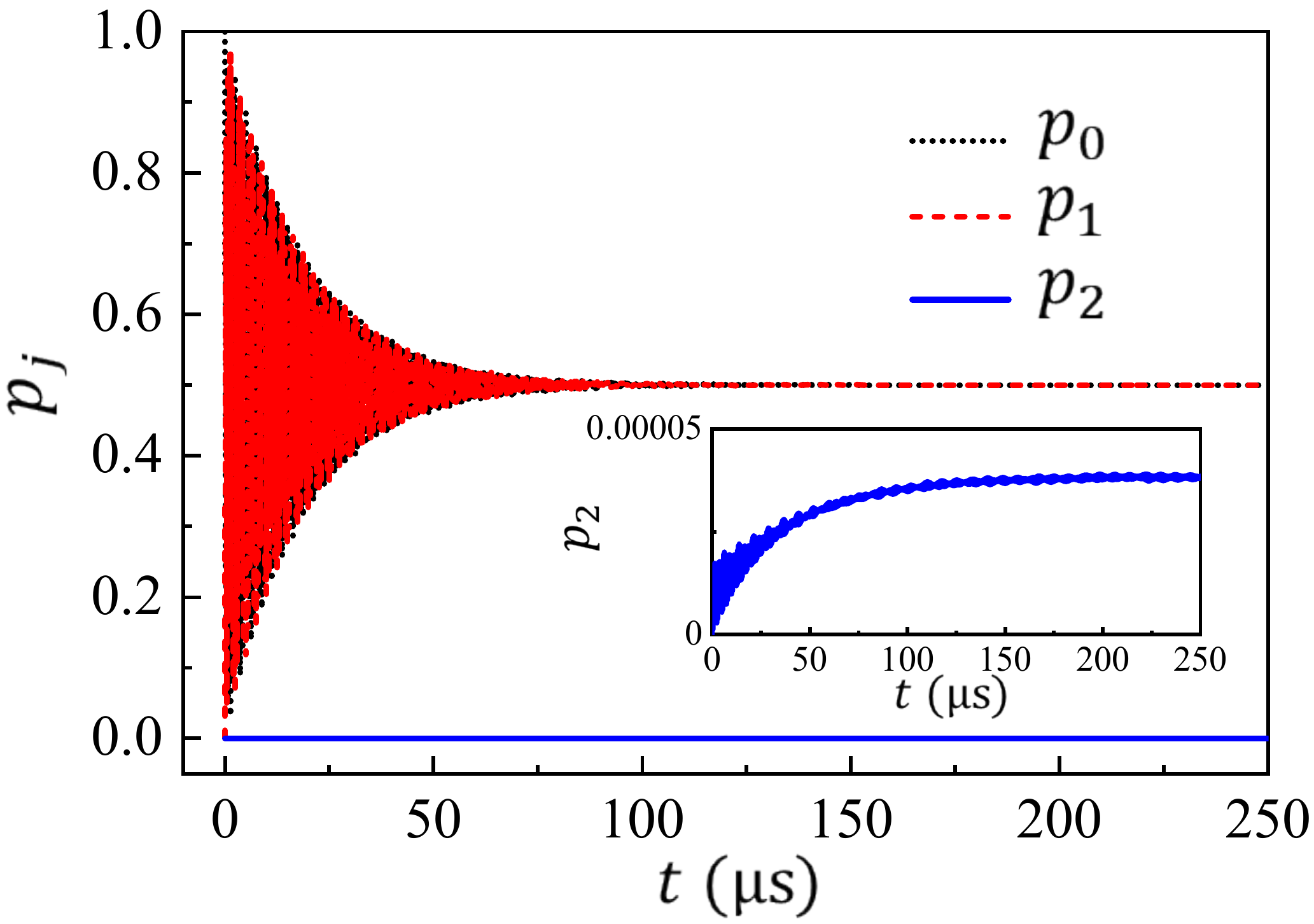}
\caption{Time evolution of the probability $p_j={\rm Tr}(\rho |j\rangle \langle j|)$ that the qutrit is in the level $|j\rangle$, obtained by numerically solving Eq.~(\ref{qutrit-master-equation}), where $j=0$ (black dotted curve), $j=1$ (red dashed curve), and $j=2$ (blue solid curve). The inset displays the enlarged view of the time evolution of $p_2$. In the numerical simulation,
we assume that the qutrit is in the ground state $|0\rangle$ at $t=0$, and other parameters are $\omega_{10}/2\pi=\omega_{d}/2\pi=5$~GHz, $\omega_{21}/2\pi=4.76$~GHz~\cite{Wang19}, $\Omega_d/2\pi=1$~MHz, $\gamma_{01}^{-1}=\gamma_{12}^{-1}=60~\mu$s, $\gamma_{02}^{-1}=150~\mu$s, $\gamma_{11}^{-1}=\gamma_{22}^{-1}=25~\mu$s, and $\gamma_{00}=3\gamma_{11}$.}
\label{fig9}
\end{figure}

\subsection{The effect of the higher levels of transmon qubits on our scheme}\label{Appendix-C2}

In the main text, we neglect the effect of the higher levels of transmon qubits on our scheme. In the following, we will give some explanations.

When a transmon qutrit is driven by a microwave field with frequency $\omega_d$, the Hamiltonian of the driven qutrit is
\begin{eqnarray}\label{original-Hamiltonian}
\mathcal{H}&=&\omega_{10}|1\rangle\langle 1|+(\omega_{10}+\omega_{21})|2\rangle\langle 2|\nonumber\\
           & &+\Omega_d\left[(|1\rangle\langle 0|+|2\rangle\langle 1|) e^{-i\omega_d t}+{\rm H.c.}\right],
\end{eqnarray}
where $|j\rangle$, $j=0,1,2$, denotes the lowest three eigenstates of the transmon circuit, $\omega_{10}$ ($\omega_{21}$) is the $|0\rangle \leftrightarrow |1\rangle$ ($|1\rangle \leftrightarrow |2\rangle$) transition frequency of the qutrit, and $\Omega_d$ is the Rabi frequency of the drive field . For the transmon qutrit, $\omega_{10}$ is larger than $\omega_{21}$, i.e., $\omega_{10}>\omega_{21}$~\cite{Wang19}. In the presence of the decoherence of the qutrit, the dynamics of the qutrit is determined by the following Born-Markovian master equation:
\begin{eqnarray}\label{qutrit-master-equation}
\frac{\partial \rho}{\partial t}&=&-i[\mathcal{H},\rho]
                                    +\gamma_{01}\mathcal{L}[|0\rangle\langle 1|]\rho+\gamma_{12}\mathcal{L}[|1\rangle\langle 2|]\rho\nonumber\\
                                 & &+\gamma_{02}\mathcal{L}[|0\rangle\langle 2|]\rho+\gamma_{00}\mathcal{L}[|0\rangle\langle 0|]\rho
                                    +\gamma_{11}\mathcal{L}[|1\rangle\langle 1|]\rho\nonumber\\
                                 & &+\gamma_{22}\mathcal{L}[|2\rangle\langle 2|]\rho,
\end{eqnarray}
where $\mathcal{L}[o]\rho=o\rho o^\dag-o^\dag o\rho/2-\rho o^\dag o/2$, with $o=|j_1 \rangle\langle j_2|$, $j_1 \leq j_2\in \{0,1,2\}$, and the coefficient $\gamma_{j_1j_2}$ is the corresponding dissipation rate. Note that in Eq.~(\ref{master-1}), we write the dephasing term as $\sigma_z\rho\sigma_z-\rho$, which is equal to $\mathcal{L}[|1\rangle\langle 1|]\rho+3\mathcal{L}[|0\rangle\langle 0|]\rho$ if we only consider the ground state $|0\rangle$ and the first-excited state $|1 \rangle$.

In our scheme, the drive field is resonant with the $|0\rangle \leftrightarrow |1\rangle$ transition (i.e., $\omega_{10}=\omega_d$) but far detuned from the $|1\rangle \leftrightarrow |2\rangle$ transition, where the detuning $|\omega_{21}-\omega_d|$ between the drive field and the $|1\rangle \leftrightarrow |2\rangle$ transition is equal to the anharmonicity $\eta=\omega_{10}-\omega_{21}$ of the transmon circuit. Experimentally, the typical anharmonicity $\eta/2\pi$ is on the order of 200-300 MHz~\cite{Wang19}, which is far larger than the optimal drive strength $\Omega_{d}^{\rm(opt)}/2\pi$ ($< 1~{\rm MHz}$; cf.~Table~\ref{Table-C2} in Appendix~\ref{Appendix-B}) in our scheme. This indicates that the second-excited state $|2 \rangle$ is barely occupied in preparing Bell states and $N$-partite $W$ states. To enhance this point, we numerically simulate the dynamics of the driven qutrit by numerically solving Eq.~(\ref{qutrit-master-equation}). As shown in Fig.~\ref{fig9}, even if the qutrit is driven, the second-excited state $|2 \rangle$ is nearly not occupied, where the probability $p_2={\rm Tr}(\rho |2\rangle \langle 2|)$ of the qutrit in the state $|2\rangle$ is always smaller than 0.00005 (i.e., $p_2<0.00005$). Thus, it is reasonable to neglect the higher levels of transmon qubits in our scheme.

\end{document}